\documentclass[aps,prl,twocolumn,amsmath,amssymb]{revtex4-1}
\usepackage{microtype}

\usepackage{braket}

\usepackage{amsmath,bm}

\usepackage{float}

\usepackage{dsfont}

\usepackage{siunitx}

\usepackage{graphicx}

\newcommand{\e}{\mathrm{e}}
\newcommand{\dd}{\mathrm{d}}

\usepackage[svgnames]{xcolor}
\usepackage{hyperref}
\hypersetup{colorlinks,linkcolor={Red},citecolor={Blue},urlcolor={Blue}}

\usepackage{booktabs}

\begin{document}

\title{Experimental tests of particle-number dependent modifications of canonical commutation relations in quantum gravity}

\date{\today}
\author{Shreya~P.~Kumar}
\email{shreya.pkumar@gmail.com}
\affiliation{Institute of Theoretical Physics and Center for Integrated Quantum Science and Technology (IQST), Albert-Einstein-Allee 11, Universit{\"a}t Ulm, 89069 Ulm, Germany}

\author{Martin~B.~Plenio}
\email{martin.plenio@uni-ulm.de}
\affiliation{Institute of Theoretical Physics and Center for Integrated Quantum Science and Technology (IQST), Albert-Einstein-Allee 11, Universit{\"a}t Ulm, 89069 Ulm, Germany}

\begin{abstract}
Models of quantum gravity imply a fundamental revision of our description of position and momentum that manifests in modifications of the canonical commutation relations. Experimental tests of such modifications remain an outstanding challenge. These corrections scale with the mass of test particles, which motivates experiments using macroscopic composite particles. Here we consider a challenge to such tests, namely that quantum gravity corrections of canonical commutation relations are expected to be suppressed with increasing number of constituent particles. Since the precise scaling of this suppression is unknown, it needs to be bounded experimentally and explicitly incorporated into rigorous analyses of quantum gravity tests. We analyse this scaling based on concrete experiments involving macroscopic pendula and provide tight bounds that exceed those of current experiments based on quantum mechanical oscillators. Furthermore, we discuss possible experiments that promise even stronger bounds thus bringing rigorous and well-controlled tests of quantum gravity closer to reality.
\end{abstract}

\maketitle

\section*{Introduction}

A minimum length scale of the order of Planck length is a feature of many models of quantum gravity (QG) that seek to unify quantum mechanics and gravitation~\cite{Garay1995a}.
But, the existence of such a length scale has so far eluded experimental verification. Direct detection of the Planck length, $10^{-35}\,\mathrm{m}$, is infeasible with
current and foreseeable technology because the effects of quantum gravity are expected to become directly relevant only at energies of the order of Planck energy,
$10^{19}\,\mathrm{GeV}$, which is 15 orders of magnitude larger than the energy scales achievable in the Large Hadron Collider today. So in order to experimentally probe quantum gravity, we must rely on indirect tests of
signatures of Planck length.

One class of indirect tests of a minimal length scale is based on observations of photon arrival times from gamma-ray bursts in distant galaxies~\cite{Amelino-Camelia1998a}.
Such experiments, however, are hard to control as they rely on a wide range of hard to verify model assumptions. These include, but are not limited to, the properties of the
models of the origins of gamma-ray bursts and the perturbations due to billions of light years of interstellar medium traversed by the gamma-rays. Furthermore, the precision
of such experiments is difficult to extend due to limitations on the distance to observable gamma-ray bursts and the maximal energy of the gamma-rays. This motivates the quest
for highly controlled table-top experiments to test for Planck scale physics \cite{Pikovski2012a,Albrecht2014a,Bawaj2015a,Bosso2017b,Kumar2018a,Bushev2019a}.

The underlying concept on which several such table-top experiments rely upon is the deformations of the canonical commutation relations of position and momentum as a consequence of a variety
of formulations of quantum gravity~\cite{Maggiore1993a,Garay1995a,Scardigli1999a,Adler1999a,Ahluwalia2000a}. These deformations are a phenomenological approach to modelling the
existence of a minimum length scale in quantum gravity and different models of quantum gravity involve different forms of modification of the commutator~\cite{Kempf1995a,Maggiore1993b,Ali2009a}.
Here we focus on a paradigmatic model~\cite{Kempf1995a}
\begin{equation}
\left[ x, p \right] = i \hbar \left( 1 + \frac{\beta_{0}}{ (M_{\mathrm{p}} c)^{2} } p^{2}\right),
\label{Eq:Beta_deformation}
\end{equation}
where $M_{\mathrm{p}}=2.176435\times 10^{-8}\, \mathrm{kg}$ and $c = 299792458 \,\mathrm{m s^{-1}}$ are the Planck mass and speed of light respectively and the variables $x$ and $p$ are the $x-$components of the position and momentum respectively. $\beta_{0}$ is a
dimensionless parameter which is expected to be of the order of unity if the minimal length scale is of the order of Planck length, but it is not fixed by
theory. Measuring or placing bounds on this parameter is thus an open experimental challenge.

One approach to bounding the value of $\beta_{0}$ is to use single particle systems, for example, using measurements of Landau levels, of the Lamb shift, or of electron tunnelling through a potential barrier~\cite{Das2008a}.
 However, the best bound obtained with these methods is $\beta_{0} < 10^{20}$, which is far from the expected $\beta_{0} \sim 1$.
To improve the bounds significantly, recent experimental proposals suggest using massive composite systems rather than elementary particles.
These experiments aim to exploit the fact that the quantum gravity signal is enhanced with larger momenta, which result from larger system mass.
Experiments and proposals in this direction include those based on the change in resonant frequency of a harmonic oscillator~\cite{Bawaj2015a,Bushev2019a,Marin2012a}, the change in broadening times of large molecular wave-packets~\cite{Villalpando2019a}, and optomechanical schemes~\cite{Pikovski2012a,Bosso2017b,Kumar2018a}.

However, the implications of using multi-particle systems to probe quantum gravity are not clear. This is because the deformations of the canonical commutation relations like Eq.~\eqref{Eq:Beta_deformation} have been derived for point particles and not for centre of mass (COM)
modes of multi-particle objects~\cite{Amelino-Camelia2013a}. 
The deformations for the COM modes are expected to decrease with the number of constituent particles in the test object~\cite{Amelino-Camelia2013a,Magueijo2003a}, but the exact expression for this suppression is not known and therefore needs to be bounded by experiment. 
Studies of the soccer-ball problem~\cite{Amelino-Camelia2011a,Hossenfelder2014a,Amelino-Camelia2017a}, which arises in a different framework of quantum gravity, also point toward a suppression of the Planck-scale corrections with number of particles.
Even if the scaling with particle number of this suppression was known, the question of what
constitutes a fundamental particle remains open. 

To address these challenges, we define a new parameter that accounts for the suppression of quantum gravity corrections with the number of constituent particles, irrespective of how these particles are defined.
Such a parameter may be obtained from theory once a full quantum description of gravity is available, and meanwhile it can be estimated or bounded from experimental observations.
We study past experiments and propose new experiments to obtain bounds on this and established parameters of quantum gravity deformations.

\section*{Results}

\subsection*{Overview}

To account for the unknown scaling
law that governs the suppression of corrections to the canonical commutation relations with the number of particles, we define the parameter $\alpha$
such that the deformation is suppressed by $N^{\alpha}$, i.e.,
\begin{equation}
\left[ x, p \right] = i \hbar \left( 1 + \frac{\beta_{0}}{ N^{\alpha} (M_{\mathrm{p}} c)^{2} } p^{2}\right)
\label{Eq:Beta_deformation_alpha}
\end{equation}
where $N$ is the number of constituent particles in the test object.
The reason for considering such a polynomial scaling with number of particles $N$ (with exponent $\alpha$) is the expected polynomial dependence in quasi-rigid macroscopic bodies, i.e., those whose centre-of-mass momenta are the sum of constituents' individual momenta~\cite{Amelino-Camelia2013a}.
Once phenomenological evidence of deformed commutator models begins to accumulate, more refined models can be considered accounting for differences in the exact nature of commutator deformation experienced by bodies with different nature of interaction.
Furthermore, we note that other models of quantum gravity could be considered, for instance, those involving non-commuting spacetime coordinates.
However, our focus is on deformed position and momentum commutators because these underlie proposed tabletop tests of quantum gravity.

In this work, we argue that the inclusion and hence estimation of $\alpha$ is essential for the rigorous interpretation of any experiment that
uses composite test objects to measure $\beta_{0}$. We propose to assess any such experiment by the exclusion area in a two-dimensional parameter
space spanned by $\alpha$ and $\beta$ and carry out such an analysis for three experiments. While the precise value of $\alpha$ is unknown, it is commonly
accepted that it needs to be positive ~\cite{Amelino-Camelia2013a,Magueijo2003a}. We shall see that the best bounds that can be calculated from
recent experiments based on micro- and nano-scale quantum harmonic oscillators \cite{Bawaj2015a,Bushev2019a} are in fact negative for $\beta_0=1$
(and in fact any $\beta_0< 10^6$).

Here we show that measured data of a macroscopic pendulum reveals the first positive
bound on $\alpha$ for any value of $\beta_{0} > 10^{-2}$ based on a careful analytical examination of the effect of deformations of the canonical
commutation relations (Eq.~\eqref{Eq:Beta_deformation_alpha}) on the time period of a pendulum. Specifically, we obtain $\alpha > 0.07$ for a
value of $\beta_{0}=1$ which is expected in various models of quantum gravity. The reason for this significant enhancement of the bound over
those obtained from micro- and nano-scale quantum harmonic oscillators can be traced back to the low achievable momenta in those experiments
which in turn lead to very weak deformations of the canonical commutation relations, for example, in the second term in Eq.~\eqref{Eq:Beta_deformation_alpha}.
We argue that our bound on $\alpha$ can be improved further by moving a pendulum to a vacuum set-up with optimised low damping suspension or
by moving to diamagnetically levitated systems which exhibit extremely low damping rates \cite{Bhattacharya2017a,Zheng2019a} on earth and promise even
better values when located in a space probe.

We begin with a study of the effects of deformed canonical commutation relations proposed in theories of quantum gravity on the time period of
a macroscopic pendulum and use these analytical results to place a bound on the parameter $\alpha$. Since a pendulum is not strictly a harmonic
oscillator, we extend the calculations of corrections to the time period of a harmonic oscillator to that of a pendulum. The time period of a
pendulum as a function of its amplitude can be determined experimentally with considerable precision. Using data from one such experiment \cite{Smith1964a}
we provide the first positive bound on $\alpha$ and moreover, suggest refined experiments for substantially improved bounds.

\subsection*{Correction to time period of pendulum}
\label{Sec:Timeperiod}
Here we calculate the corrections to the time period of a pendulum due to quantum gravity deformations of the canonical commutation relations so that
it can then be compared against experimental data. We consider a pendulum of mass $m$ and length $L$. Its Hamiltonian is
\begin{equation}
H = \frac{p^{2}}{ 2 m} \cos^{2} \theta - m g L \cos \theta
\label{Eq:H}
\end{equation}
where $p$ is the generalised momentum conjugate to $x$ and $\theta$ is the angle between the radius vector and the vertical, i.e., $\theta = \sin^{-1}\left(x/L\right)$.
The time period can be obtained from the Hamiltonian in two ways. Here, we take the approach that defines a new momentum operator which satisfies the standard
canonical commutation relations and hence the standard Heisenberg equations of motion. In this approach the Hamiltonian is modified. In an alternative, equivalent,
approach the Hamiltonian can be left unchanged while the equations of motion are modified due to the deformed commutator~\cite{Nozari2005b}. Here we follow the first
method due to ease of calculation.

For illustrative purposes, we perform the calculations using classical mechanics but later show that the results hold even by performing quantum calculations with deformed commutators.
Here, we deform the standard Poisson brackets in analogy to the deformation of the canonical commutation relation due to quantum gravity~\cite{Benczik2002a,Nozari2008a,Pedram2012a}, i.e.,
\begin{equation}
\left\{ x, p \right\} = 1 + \beta p^{2}
\end{equation}
where
\begin{equation}
\beta = \frac{\beta_{0}}{N^{\alpha} (M_{\mathrm{p}} c)^{2}}.
\label{Eq:beta}
\end{equation}
To ensure that the equations of motion are unchanged, a new momentum operator $\tilde{p}$ is defined such that we recover the standard Poisson bracket, i.e.,
\begin{equation}
\left\{ x,  \tilde{p} \right\} = 1.
\end{equation}
Without the small momentum assumption of Refs.~\cite{Bawaj2015a,Bushev2019a}, we find that $\tilde{p}$ is related to $p$ as
\begin{equation}
\tilde{p}= \frac{ \tan^{-1} \left(\sqrt{\beta} p \right) }{\sqrt{\beta}}.
\label{Eq:NewP}
\end{equation}
Writing the Hamiltonian in terms of the new momentum operator $\tilde{p}$, we note that it is modified to
\begin{equation}
H = \frac{1}{ 2 m \beta} \tan^{2} \left(\sqrt{\beta} \tilde{p} \right) \cos^{2} \theta  - m g L \cos \theta.
\label{Eq:ModH}
\end{equation}
This modification of the Hamiltonian compared to Eq.~\eqref{Eq:H} ensures that the equations of motion remain unchanged with respect to $x$ and $\tilde{p}$, i.e.,
\begin{equation}
\dot{x} = \frac{\partial H}{\partial \tilde{p}}
= \frac{\cos^{2} \theta}{ m \sqrt{\beta}} \tan \left(\sqrt{\beta} \tilde{p} \right)\sec^{2} \left(\sqrt{\beta} \tilde{p} \right).
\label{Eq:xt_pendulum}
\end{equation}
Separating the variables and integrating over half a time period as detailed in the Methods, we obtain the time period for small amplitudes $A$ and small $\beta$ as approximately
\begin{equation}
T_{2 \pi}  \approx 2 \pi \sqrt{\frac{L}{g}}
\left( 1 + \frac{A^{2}}{16 L^{2}} - \frac{\beta_{0} m^{2} g }{2 N^{\alpha} (M_{\mathrm{p}} c)^{2} L} A^{2} \right).
\label{Eq:TP_approx}
\end{equation}
Here, we have calculated the correction to the time period to first order in the quantum gravity parameter $\beta$. 
We argue that if $\beta$ were not, in fact, small, then a much larger deviation in the time period would be observed.
In such a case, more detailed calculations would be required to obtain accurate bounds.
We show that the time period of a pendulum obtained from experimental data in the next section corroborates this assumption.

In the Methods section, we show that the Poisson equation approach and a fully quantum mechanical calculation yield the same result for the time period of a harmonic
oscillator. This shows that the calculated expression for the time period (Eq.~\eqref{Eq:TP_approx}) is not merely a
consequence of using classical dynamics but extends to the quantum regime and thus provides further evidence for the correctness of our approach.
The detailed calculations are presented in the Methods section, but we summarise the arguments here. To perform these calculations, we start
with the eigenvalues and eigenfunctions of the quantum harmonic oscillator derived in Ref.~\cite{Kempf1995a}. Since ladder operators are
defined differently due to commutator deformation, we use a generalised Heisenberg algebra~\cite{Pedram2013a} to find the action of the ladder
operators on the eigenstates. Using this algebra, we derive the expressions for the position and momentum operators in terms of the ladder operators.
With this, the operators are well defined. In order to choose the most classical pure state in our calculations, we choose a definition of coherent
states, the Gazeau-Klauder states~\cite{Gazeau1999a}, such that the states remain coherent states during the evolution under this Hamiltonian.
Calculating the expectation value of the position operator with respect to these Gazeau-Klauder coherent states, we recover the classical calculations.

As a side remark we note that these matching results also connects two different approaches to studying deformed commutators, namely modifying the Poisson bracket~\cite{Benczik2002a,Nozari2008a,Pedram2012a} and modifying the commutator~\cite{Kempf1995a,Ali2011a,Brau1999a}. 
These two approaches have so far been thought to be separate~\cite{Scardigli2015a}.

\subsection*{Bounds on QG parameters from experimental data}
\label{Sec:Exp}

Having calculated the correction to the time period of a pendulum, we use experimental data from precise measurements of the time period of a pendulum to place bounds on the quantum gravity parameters $\alpha$ and $\beta_{0}$ using Eq.~\eqref{Eq:TP_approx}.

To this end, we consider an experiment~\cite{Smith1964a} which measures the time-period of a pendulum as a function of its amplitude. 
The experimental data is represented in Fig.(3) of \cite{Smith1964a}, which plots the measured time-period as a function of the square of the amplitude. 
The figure contains data from two experiments: one using a conventional suspension of the pendulum and another using a cycloidal suspension. 
In this manuscript, we consider only the conventional suspension because such an experiment can be modelled by the calculations shown in the previous section.

We extracted the data of the experiment with the conventional suspension from Fig.(3) to calculate the slope and intercept. 
From Eq.~\eqref{Eq:TP_approx}, we see that the intercept $T_{0}$ of this line is $2 \pi \sqrt{\frac{L}{g}} $ and the slope is $ 2 \pi \sqrt{\frac{L}{g}} \left( \frac{1}{16 L^{2}} - \frac{\beta_{0} m^{2} g }{2 N^{\alpha} (M_{\mathrm{p}} c)^{2} L} \right).$ 
Since the expression for the slope of this plot includes corrections from quantum gravity, we calculate the slope from the extracted data from Fig.(3) to compare with theory. 
This data is reported and the method of extraction of the data is detailed in Sec~\ref{Sec:Data}.

We use the data extracted to perform a linear fit. 
The reported error of measurements is $2\%$ in amplitude measurement and $3 \times 10^{-5}~\mathrm{s}$ in time-period measurement~\cite{Smith1964a}.
However, additional error arising from the datapoint extraction have to be accounted for. 
For this, we use the size of the markers as a conservative estimate.
This leads to an error of $5 \times 10^{-3}~\mathrm{m}^{2}$ in the square of the amplitude and  $10^{-4}~\mathrm{s}$ in the time period measurements.

We use the extracted data and account for both the sources of errors to perform a linear fit using the orthogonal distance regression method~\cite{Boggs1992a}. 
This fit yields a reduced $\chi^{2}$ value of $0.07$, which indicates that the extracted data points do indeed agree with the linear fit.
We note that the current work is only a blueprint that uses experimental data from 1964 and such an analysis can be repeated with improved experiments to determine tighter and more precise bounds.

From this fit we obtain the value of the intercept
\begin{equation}
T_{0} = \SI{3.4730 \pm 0.0001}{s}.
\end{equation}
Using this value and following the analysis of Ref.~\cite{Smith1964a}, we can precisely infer the effective length of the pendulum to be
\begin{equation}
L = g \left( \frac{T_0}{2 \pi} \right)^{2}= \SI{2.9954 \pm 0.0002}{m}.
\end{equation}
In this calculation, the local value of acceleration due to gravity $g = \SI{9.80393}{m.s^{-2}}$ has been used~\cite{Smith1964a}.
The pendulum is an iron cylinder of radius $\SI{2.54}{cm}$ and height $\SI{5.08}{cm}$ and a mass of approximately $\SI{1.22}{kg}$.

Using the obtained value of the length $L$ and mass $m$, the slope is numerically evaluated to be
\begin{equation}
2 \pi \sqrt{\frac{L}{g}} \left( \frac{1}{16 L^{2}} - \frac{\beta_{0} m^{2} g }{2 N^{\alpha} (M_{\mathrm{p}} c)^{2} L} \right) = 0.0242 - 0.197 \frac{\beta_{0}}{N^{\alpha}}.
\label{Eq:SlopeTheory}
\end{equation}
The uncertainty in the obtained slope due to uncertainty in the derived length of the pendulum is $9 \times 10^{-5}$, which is small enough to be ignored in the following calculations.

From the linear fit of the data, the slope of the line is obtained to be $0.0232 \pm 0.0012$ ($95\%$ confidence interval).
We see that for the fit and Eq.~\eqref{Eq:SlopeTheory} to be consistent, we obtain $ -0.0012 < \frac{\beta_{0}}{N^{\alpha}} < 0.011$.
For $\beta_{0}$ positive,
\begin{equation}
\beta_{0}N^{-\alpha} < 10^{-2}.
\end{equation}

For the determination of $N$ we assume that the nucleons form the elementary particles which leads to $N = 7.32 \times 10^{26}$
and therefore we obtain $\alpha > 0.07$ for $\beta_{0}=1$. Note that the bound on $\alpha$ is quite insensitive to the precise number of nucleons.

We note that such tight bounds were possible in our work as compared to those of previous works because the nonlinearities in a pendulum can be computed precisely. 
In other systems, the intrinsic nonlinearities are unknown and therefore contribute to larger bounds on quantum gravity parameters.

These bounds can therefore be made tighter by including the effect of dissipation on the change in frequency of a pendulum.
We propose that future experiments measure the amount of dissipation and include its effect on the slope and intercept of the $T$ versus $A^{2}$ plot in order for the bounds to be more precise. 
We note that dissipation is not a feature of classical systems alone and quantum systems such as those proposed for other tests also couple to the environment, and this dissipation should be considered as well.

\begin{figure}[h]
\centering
\includegraphics[width=\columnwidth]{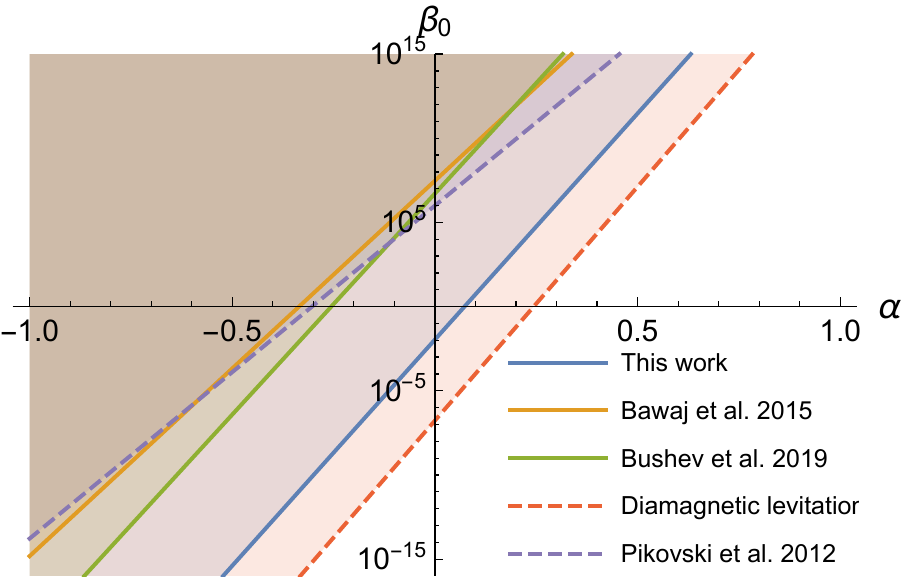}
\caption{\textbf{Excluded regions of parameter space from different experiments.} 
This figure depicts the values of the two quantum gravity parameters $\beta_{0}, \alpha$ that are excluded based on experimental observations. 
Solid lines represent bounds obtained from experimental data and dashed lines represent expected bounds from proposed experiments.
The shaded areas represent the region excluded by these experiments.
The present work based on \cite{Smith1964a} provides the largest excluded region of parameters which, in particular, excludes the key point $\beta_{0}=1, \alpha = 0$, thereby showing that suppression of quantum gravity deformations should be accounted for if $\beta_{0} \sim 1$ as expected from quantum gravity models.
The proposal to use massive levitated diamagnetic objects described in the text promises significant improvement in bounds if the optimistic parameters required for such an experiment can be obtained.
\label{Fig:Scaling}}
\end{figure}

Recent experiment using oscillators in the quantum regime have been used to provide bounds on $\beta_0$ under the assumption that
$\alpha=0$. We argue however that any test of consequences of deformed canonical commutation relations due to quantum gravity need
to account for both $\alpha$ and $\beta_{0}$.
The best bound on $\alpha$ from the experiments of Ref.~\cite{Bawaj2015a} is $\alpha > -0.33$ for $\beta_{0}=1$. Similarly, from Ref.~\cite{Bushev2019a}
we obtain $\alpha > -0.25$. Note that these bounds are significantly worse than those obtained in the
present work using the data from \cite{Smith1964a}. In Fig.~\ref{Fig:Scaling} we present the parameter ranges that have been excluded
in the $\alpha,\beta_{0}$-plane in the three experiments discussed here.

\subsection*{Diamagnetic levitation for enhanced tests of QG}
In order to explore how far we can possibly bound the $\alpha$-parameter, here we propose an experiment that relies on the precise measurement
of the oscillation frequency of a diamagnetic levitated particle to obtain enhanced bounds on the quantum gravity parameters.
In an experiment on a space probe, such as LISA pathfinder, one could imagine to levitate a particle
in a uniform magnetic field gradient. In this case the frequency of oscillation would be~\cite{Pedernales2020a}
\begin{equation}
    \omega = \sqrt{\frac{1}{\rho \mu_0} \chi_{\mathrm{v}}  \left(\frac{dB}{dx}\right)^2}
\end{equation}
where $\rho$ is the mass density of the object, $\mu_0$ is the vacuum permeability, $\chi_{\mathrm{v}}$ is the
magnetic volume susceptibility of the material and we assume a constant magnetic field gradient.
The change in frequency resulting from deformed commutators can be obtained from Eq.~\eqref{Eq:TP_approx} and is given by
\begin{equation}
\Delta \omega = \frac{\beta_{0} m^{2} \omega^{3} A^{2}}{2 N^{\alpha} (M_{\mathrm{p}} c)^{2}} .
\end{equation}
The oscillation frequency of levitated objects can be measured very precisely due to very low damping rates.
Here we assume, optimistically, that the damping rate is the only source of error in frequency measurement.
At low pressures of $266 \times 10^{-10}\,\mathrm{Pa}$, the damping rate is expected to be $\gamma=1.2 \times 10^{-7}
\, \mathrm{Hz}$~\cite{Epstein1924a,Slezak2018a}. If, in the experiment, no deviation from the expected frequency
is observed, then $\Delta \omega \lesssim \gamma/\sqrt{N_{\mathrm{m}}}$ where $N_{\mathrm{m}}$ is the number of measurements taken.

We calculate the bounds that one would obtain if such an experiment can be performed. We consider optimistic parameters
of a gold sphere of diameter $\SI{10}{cm}$ that is levitated in a uniform magnetic field gradient of $10^{3}\,\mathrm{T/m}$
that is initially displaced with an amplitude of  $\SI{10}{cm}$. The density and magnetic volume susceptibility of gold are
$\rho = 19300\,\mathrm{kg/m^{3}}$ and $\chi_{\mathrm{v}} = 3.287 \times 10^{-5}$~\cite{Dupree1973a} respectively. 
This leads to the frequency of oscillations $\omega = 36.71$ Hz, from which we estimate that for $\beta_{0}=1$, we obtain $\alpha>0.35$ for a single measurement $N_{\mathrm{m}}=1$. 
However, we realise that such estimates are based on optimistic assessments of all the involved parameters.
More realistically, the error in the frequency might not limited by damping alone and the different parameters feasible in a single experiment might be somewhat suboptimal once all components are integrated together.
Thus, we can conservatively assume that the actual precision in frequency is three orders of magnitude worse than the optimistic value suggested above, i.e., $\Delta \omega \lesssim 10^{-4}$ Hz, we obtain the bounds $\alpha > 0.24$ for $\beta_{0} = 1$. 

The region of parameter space that can be excluded from such an experiment is shown in Fig.~\ref{Fig:Scaling}. This bound may be improved by performing the experiment in space, where
the pressure is about 2000 times lower which leads to a further reduction in $\gamma$.

\subsection*{Using optomechanical tests to bound $\alpha$}

Experiments to place bounds on quantum gravity parameters are not restricted to the framework of measuring the change in frequency of oscillators.
Here we show that other schemes can also be used to placed stringent bounds on $\alpha$ if appropriate initial states are considered.

Specifically, we consider the optomechanical scheme of a light pulse striking a mechanical resonator repeatedly and acquiring a phase that depends
on quantum gravity corrections to the canonical commutation relations~\cite{Pikovski2012a,Bosso2017b,Kumar2018a}. In this scheme, the mechanical
resonator is initially in a thermal state very close to the ground state.
After interacting with the resonator, the field of the light pulse is~\cite{Pikovski2012a}
\begin{equation}
\braket{a_\ell} \approx  \xi
\e^{-i 2 \lambda^{2} N_{p} -i \left(4/3 \right) \beta \hbar m \omega \lambda^4 N_{p}^3 }
\end{equation}
where $\xi$ is the amplitude of the coherent state $\ket{\xi}$ of light, $N_{p}$ is the mean photon number in the light pulse and $\lambda$ is the optomechanical interaction strength. As before, $m$ and $\omega$ are the mass and frequency of the oscillator.
We see that the phase acquired by light has a term dependent on $\beta$, which originates from the additional contribution from quantum gravity.
Using the experimental parameters of Ref.~\cite{Pikovski2012a} and the error analysis of Ref.~\cite{Kumar2018a}, and assuming that experiment
returns a null result, we obtain the bound $\beta_{0} N^{-\alpha} < 10^{6}$ which leads to $\alpha > -0.3$ for $\beta_{0}=1$. The excluded
parameter range is presented in Fig.~\ref{Fig:Scaling}.

This bound can be improved if, instead of the ground state, the resonator is initially in a coherent state with a large enough momentum.
In this case, the output field is given by
\begin{align}
\begin{split}
\braket{a_\ell} \approx& \, \xi
\e^{-i 2 \lambda^{2} N_{p} -i \left(4/3 \right) \beta \hbar m \omega \lambda^4 N_{p}^3 + i \beta \lambda^{2} N_{p}  2 \braket{p}^{2}}
\end{split}
\end{align}
for the initial state of the oscillator in a coherent state with mean initial momentum $\braket{p}$.

To understand better the mass dependence of the quantum gravity terms, we consider the mass dependence of $\lambda$ using an established model~\cite{Pikovski2012a}
\begin{equation}
\lambda = \frac{4 \mathcal{F}}{\lambda_{\mathrm{L}}} \sqrt{\frac{\hbar}{m \omega}}
\end{equation}
where $\mathcal{F}$ is the finesse of the cavity and $\lambda_{\mathrm{L}}$ is the wavelength of light used. 
With this scaling of $\lambda$ with the mass $m$, we see that the first negative term scales as $\frac{1}{m}$ while the second one scales as $m$. 
Hence, we see that the extra term arising from a non-zero initial momentum can, in principle, be made large by choosing a more massive oscillator. 
Making the optimistic assumption that all other parameters remain the same but we increase the mass to $10^{-3}\, \mathrm{kg}$ a positive bound on $\alpha$ would be achieved. 
While such a large mass is likely to reduce the optomechanical coupling it nevertheless suggests that the measurement of the phase of the output light for macroscopic systems may provide another method to obtain a good bound on the deformation parameters.

\section*{Discussion}
Quantum gravity suggests corrections to the canonical commutation relations that are proportional
to a parameter $\beta_{0}$.
This parameter is expected to be of order of unity if physics exhibits a minimum length of order of the Planck length
but is also expected to scale as $N^{-\alpha}$ where $N$ is the number of constituent particles of the test mass -- a
consequence of the soccer ball problem of quantum gravity. We strongly argue that any test of such physics needs to account
for both parameters $\alpha$ \textit{and} $\beta_{0}$ in its analysis.
We perform an analysis of several quantum regime experiments in
those terms to show that they cannot provide positive bounds on $\alpha$ while we find that a macroscopic pendulum can
provide the first positive bound on $\alpha$ assuming $\beta_{0} = 1$. This shows that the suppressions with the number
of particles cannot be ignored in tests of quantum gravity and that entering the deep quantum regime is not essential
for the observation of quantum gravity corrections to physical dynamics. 

Our results pertaining to the time period are derived using the method of deformed Poisson brackets~\cite{Benczik2002a,Nozari2008a,Pedram2012a}. We
also put this method on a more rigorous footing by connecting it to the calculations based on deformed commutators~\cite{Kempf1995a,Ali2011a,Brau1999a}
and show that the two results match. Hence our work connects these two approaches that have thus far been considered independent~\cite{Scardigli2015a}.
Finally, we show that the suppression of quantum gravity deformations is not just restricted to this one framework of oscillator frequency measurement.
For instance, we consider the optomechanical system of Refs.~\cite{Pikovski2012a,Bosso2017b,Kumar2018a} and verify that broadly analogous considerations hold.
We discuss possible advanced experimental designs and the parameter requirements to allow for entering tests in the $\alpha>1$ regime that is suggested by various models of quantum gravity.

\section*{Methods}
\label{Sec:Methods}

\subsection*{Time period of pendulum: detailed calculations}
In this section, we obtain the expression for the time period of a pendulum using classical mechanics.
This is done by solving the equation of motion for the modified Hamiltonian obtained in the calculation of the correction to time period of pendulum in the Results section.

From the expression for the energy in Eq.~\eqref{Eq:ModH}, the total energy in the system is $E = -m g L \cos \phi$, where $\phi$ is the angular amplitude.
The redefined momentum $\tilde{p}$~\eqref{Eq:NewP} can be expressed in terms of the angular displacement $\theta$ and amplitude $\phi$ as
\begin{equation}
\tilde{p} = - \frac{1}{\sqrt{\beta}} \tan^{-1} \left( \sqrt{ 2 m^{2} g L f(\theta) \beta} \right)
\end{equation}
where
\begin{equation}
f(\theta) :=  \frac{ \cos \theta - \cos \phi }{\cos^{2} \theta}.
\end{equation}

The equation of motion Eq.~\eqref{Eq:xt_pendulum} can be rewritten in terms of $\theta$ as
\begin{equation}
\dot{\theta} L \cos \theta = - \frac{\cos^{2} \theta}{ m \sqrt{\beta}}  \sqrt{ 2 m^{2} g L f(\theta) \beta} \left(1+ 2 m^{2} g L f(\theta) \beta \right)
\end{equation}
and simplified to
\begin{equation}
\dot{\theta}   = - \cos \theta \sqrt{ 2 \frac{g}{L} f(\theta)} \left(1+ 2 m^{2} g L f(\theta) \beta \right).
\label{Eq:ClassicalDE}
\end{equation}
Separating variables and integrating over half a cycle, we obtain the time period for half an oscillation
\begin{equation}
\frac{T_{2\pi}}{2} = \int_{-\phi}^{\phi} \dd \theta  \frac{1}{ \cos \theta \sqrt{ 2 \frac{g}{L} f(\theta)} \left(1+  2 m^{2} g L f(\theta) \beta \right)} .
\end{equation}
Since $\beta \ll 1$, as can be numerically verified from the above equation, the expression can be simplified to
\begin{align}
\begin{split}
T_{2 \pi} \approx \sqrt{\frac{2 L}{g}}  \int_{-\phi}^{\phi} \dd \theta & \left\{ \frac{1}{ \sqrt{ \cos \theta - \cos \phi} } \right.
\\
 & \left. - \frac{  \beta 2 m^{2} g L \sqrt{ \cos \theta - \cos \phi} }{  \cos^{2} \theta} \right\}.
\end{split}
\end{align}
Furthermore, for small amplitudes, we can approximate the time-period to
\begin{equation}
T_{2 \pi}  \approx 2 \pi \sqrt{\frac{L}{g}}
\left( 1 + \frac{\phi^{2}}{16} - \frac{\beta}{2} m^{2} g L \phi^{2} \right)
\label{Eq:PedulumApprox}
\end{equation}
and in terms of the amplitude $A$, where $A = \phi L$, it can be expressed as
\begin{equation}
T_{2 \pi}  \approx 2 \pi \sqrt{\frac{L}{g}}
\left( 1 + \frac{A^{2}}{16 L^{2}} - \frac{\beta_{0} m^{2} g }{2 N^{\alpha} (M_{\mathrm{p}} c)^{2} L} A^{2} \right)
\end{equation}
which is the expression that has been used to compare with experiments.

\subsection*{Rigorous calculations using deformed commutators}

In this section, we perform detailed calculations in a quantum mechanical framework to find the time-dependent position of a harmonic oscillator modified by quantum gravity.
To do so, we start by using the energy eigenvalues and eigenstates derived quantum-mechanically in Refs.~\cite{Kempf1995a,Chang2002a} for deformed commutators.
We then choose appropriate definitions of the ladder operators and derive the position operator in terms of these ladder operators.
The definition of a coherent state is chosen such that the state remains invariant under evolution of the Hamiltonian.

In the following calculations, we work in the momentum basis where the operators $\hat{x}$ and $\hat{p}$ are defined by their action on the momentum wave-functions as
\begin{align}
\hat{x} \psi(p)&= i \hbar (1 + \beta p^{2}) \frac{\partial \psi(p)}{\partial p}
\label{Eq:x}
\\
\hat{p}\psi(p) &= p\psi(p).
\label{Eq:p}
\end{align}

Solving the Schr\"odinger equation $\hat{H} \psi_{n}(p) = E_{n} \psi_{n}(p)$ with Hamiltonian
\begin{equation}
\hat{H} = -\frac{\hbar^{2} m \omega^{2} }{2}  \left( (1 + \beta p^{2}) \frac{\partial}{\partial p} \right)^{2} + \frac{p^{2}}{2 m},
\end{equation}
the energy eigenvalues are found to be~\cite{Kempf1995a}
\begin{equation}
E_{n} = \hbar \omega \left( n + \frac{1}{2} \right) \left( \sqrt{1 + \frac{1}{16 r} } + \frac{1}{4 \sqrt{r} }\right) + \frac{\hbar \omega}{4 \sqrt{r}} n^{2}
\label{Eq:eigenval}
\end{equation}
for $1/r = (2 \beta m \hbar \omega)^{2}$.
The eigenfunctions in the momentum basis are~\cite{Chang2002a}
\begin{align}
\begin{split}
\psi_{n}(p) &= (-i)^{n} 2^{\lambda} \Gamma(\lambda) \sqrt{ \frac{n! (n + \lambda) \sqrt{\beta} }{ 2 \pi \Gamma(n + 2 \lambda)} } (1 - s^{2})^{\lambda/2} C_{n}^{\lambda}(s)
\\
&=: z_{n} (1 - s^{2})^{\lambda/2} C_{n}^{\lambda}(s)
\end{split}
\label{Eq:eigenfunc}
\end{align}
where $C_{n}^{\lambda}(s)$ are Gegenbauer polynomials and
\begin{align}
s &= \frac{\sqrt{\beta} p}{\sqrt{1 + \beta p^{2}}} \\
\lambda &= \frac{1}{2} + \sqrt{\frac{1}{4} + \frac{1}{(m \hbar \omega \beta)^{2}} }.
\end{align}
The phase $(-i)^{n}$ has been introduced here so that in the limit $\beta \to 0$, we recover results from quantum mechanics, namely $\hat{a} = \sqrt{\frac{m \omega}{2 \hbar}} \hat{x} + \frac{i}{\sqrt{2 m \hbar \omega}} \hat{p}$.

In the rest of the calculations, we work in the Fock basis $\left\{ \ket{n} \right\}$, using the eigenvalues given by Eq.~\eqref{Eq:eigenval} and the eigenfunctions $\psi_{n}(p) = \braket{p|n}$ given by Eq.~\eqref{Eq:eigenfunc}.
The number operator $\hat{n}$ is defined such that $\hat{n} \ket{n} = n \ket{n}$.

Next we describe the Generalised Heisenberg algebra that is used in the subsequent calculations of the position and momentum operators.
The algebra of the ladder operators should be modified to account for the deformed commutators.
Using the version of generalised Heisenberg algebra used in Ref.~\cite{Pedram2013a}, we find the action of the annihilation operator on an energy eigenstate to be
\begin{equation}
\hat{a} \ket{n} = \sqrt{n \left( 1 + \nu + \nu n \right)} \ket{n-1}
\label{Eq:Annihilation}
\end{equation}
for $\nu = \beta m \hbar \omega / 2$.
The number operator is related to the ladder operators as
\begin{equation}
\hat{a}^{\dag} \hat{a} = \hat{n} \left( 1 + \nu + \nu \hat{n} \right).
\end{equation}
Also, the commutator is derived to be
\begin{equation}
\left[ \hat{a}, \hat{a}^{\dag} \right] \approx 1 + 2 \nu (1 +  \hat{a}^{\dag} \hat{a})
\end{equation}
to first order in $\beta$.

Now we are ready to derive the expression for the position and momentum operators in terms of the ladder operators.
These calculations will eventually enable the calculation of the trajectory of the oscillator.
The relationship between $\hat{x}$ and $\hat{a}$ is obtained by starting with the equation Eq.\eqref{Eq:Annihilation} in the momentum basis, i.e,
\begin{equation}
\hat{a} \psi_{n}(p) =  \sqrt{n \left( 1 + \nu + \nu n \right)} \psi_{n-1}(p)
\end{equation}
and using recursion relations of the Gegenbauer polynomials to express $\psi_{n-1}(p)$ in terms of $\psi_{n}(p)$.
This gives us the operator $\hat{a}$ in the momentum basis which can be expressed in terms of the operators $\hat{x}$ and $\hat{p}$.
This relation is then inverted to obtain $\hat{x}$ in terms of the ladder operators.

Using the following recursion relations
\begin{align}
(n + 2 \lambda) C_{n}^{\lambda}(s) &= \frac{ \dd}{\dd s} C_{n+1}^{\lambda}(s) - s  \frac{ \dd}{\dd s} C_{n}^{\lambda}(s)
\\
(1 - s^{2}) \frac{ \dd}{\dd s} C_{n}^{\lambda}(s) &= (n + 2 \lambda) s C_{n}^{\lambda}(s) - (n + 1) C_{n+1}^{\lambda}(s)
\end{align}
of Gegenbauer polynomials, we obtain the action of the annihilation operators on the wavefunction to be
\begin{align}
\begin{split}
\hat{a} \psi_{n} =& \, i \sqrt{\frac{  \left( 1 + \nu + \nu n \right)  (n + \lambda -1) (\lambda + n) \beta}{(n + 2 \lambda -1) (1 + \beta p^{2})}}
\times  \\
& \left\{ \frac{1 + \beta p^{2}}{\beta (\lambda + n) } \frac{ \dd }{\dd p}
+p \right\} \psi_{n}.
\label{Eq:aExact}
\end{split}
\end{align}
Since the $\psi_{n}$ form a complete basis~\cite{Kempf1995a}, Eq.~\eqref{Eq:aExact} can be written in operator form in the limit of $\beta \ll 1$ using the definitions of the position and momentum operators,~\eqref{Eq:x} and~\eqref{Eq:p}, to obtain
\begin{align}
\begin{split}
\hat{a}  =&\,  \sqrt{\frac{m \omega}{2 \hbar}} \left[ \hat{x} - \beta \left\{ \frac{ 1}{2} \hat{p}^{2} \hat{x} + \frac{m \hbar \omega}{4}  \hat{x} \right\} \right]
 \\
&+ \frac{i}{\sqrt{2 m \hbar \omega}} \left[ \hat{p}
 + \frac{\beta}{4} \left\{  - 2 \hat{p}^{3}  + \hbar m \omega \hat{p} (1 + 4 \hat{n})  \right\} \right].
\end{split}
\end{align}
We invert this relation to find $\hat{x}$
\begin{align}
\begin{split}
\hat{x} =& \, \sqrt{\frac{\hbar}{2 m \omega}} \left( \hat{a} + \hat{a}^{\dag} \right)
 \\
&  + \frac{\beta}{4} \sqrt{\frac{\hbar^{3}  m \omega }{2}}   \left(  \hat{a}^{\dag} \hat{a}^{2} +  \hat{a}^{\dag 2} \hat{a}  -\hat{a}^{3} - \hat{a}^{\dag 3}  \right)
\end{split}
\label{Eq:xtoa}
\end{align}
and similarly $\hat{p}$
\begin{equation}
\begin{split}
\hat{p} = & \, i \sqrt{\frac{\hbar m \omega}{2}} \left( \hat{a}^{\dag} - \hat{a} \right)
\\
& + i \beta \frac{(\hbar m \omega)^{3/2} }{4 \sqrt{2}}  \left(  \hat{a}^{\dag} \hat{a}^{2} -  \hat{a}^{\dag 2} \hat{a}  + \hat{a}^{3} - \hat{a}^{\dag 3} + 2 \hat{a} - 2 \hat{a}^{\dag}  \right)
\label{Eq:ptoa}
\end{split}
\end{equation}
in terms of the ladder operators.

Next we move to the definition of the generalised coherent state.
To define the most classical pure state, a natural choice is a coherent state, but due to the Hamiltonian being modified from deformed commutators, our usual definition of coherent states no longer hold because the coherent state does not remain one after evolution under this Hamiltonian. 
Hence, here we introduce generalised coherent states that are suited for this modified Hamiltonian.
These states are the Gazeau-Klauder states which were introduced in Ref.~\cite{Gazeau1999a}.
Here, we describe them in detail, closely following the details in Ref.~\cite{Gazeau1999a}.

Since a coherent state $\ket{\alpha}$ is parametrised by one complex number, we generalise it slightly by considering states parametrised by two real parameters $\ket{J, \gamma}$.
To ensure that $\ket{J, \gamma}$ behaves like a classical state, we demand that it satisfies the following conditions with respect to a given Hamiltonian $H$:
\begin{enumerate}
\item
The continuity condition
\begin{equation}
(J', \gamma') \rightarrow (J, \gamma)  \implies \ket{J', \gamma'} \rightarrow \ket{J, \gamma}
\end{equation}
\item
Resolution of identity
\begin{equation}
\int \dd \mu (J, \gamma) \ket{J, \gamma} \bra{J, \gamma}  = \mathds{1} 
\end{equation}
\item
Temporal stability such that the time-evolved state is always a generalised coherent state
\begin{equation}
\e^{-i H t/\hbar} \ket{J, \gamma} = \ket{J, \gamma + \omega t}
\label{Eq:TimeEvol}
\end{equation}
\item
The energy of the state only depends on $J$
\begin{equation}
\braket{J, \gamma | H | J, \gamma} = \hbar \omega J.
\label{Eq:ActionId}
\end{equation}
\end{enumerate}
 
The conditions in Eqs.~\eqref{Eq:TimeEvol} and~\eqref{Eq:ActionId} are defined with respect to a Hamiltonian, and so coherent states do not satisfy them with respect to a modified Hamiltonian and we need these generalised states.

If the eigenvalues and eigenstates of the Hamiltonian are defined such that 
\begin{equation}
H \ket{n} = \hbar \omega e_{n }\ket{n},
\end{equation}
we can verify that the definition  of the generalised coherent state 
\begin{equation}
\ket{J, \gamma} = \frac{1}{N(J)} \sum_{n} \frac{J^{n/2} \e^{-i \gamma e_{n}}}{\sqrt{\rho_{n}}} \ket{n}
\label{Eq:Gazeau-Klauder}
\end{equation}
where
\begin{equation}
\begin{split}
\rho_{n} &= \prod_{k=1}^{n} e_{k}\\
N(J)^{2} &= \sum_{n} \frac{J^{n}}{\rho_{n}} 
\end{split}
\end{equation}
satisfies all the above properties. 

As an example, we consider the harmonic oscillator Hamiltonian.
Here, the eigenvalues are given by $H \ket{n} = \hbar \omega n \ket{n}$ (after ignoring the constants) and therefore in this case, $e_{n} = n$.
Therefore, from the definitions of $\rho_{n}$ and $N(J)$, we see that the Gazeau-Klauder state is
\begin{equation}
\ket{J, \gamma} = \e^{-J/2} \sum_{n} \frac{J^{n/2} \e^{-i \gamma n}}{\sqrt{n!}} \ket{n},
\end{equation}
which is exactly the definition of a coherent state $\ket{\xi}$ if we define $\xi = \sqrt{J} \e^{-i \gamma}$.
Thus we see that the Gazeau-Klauder state reduces to the coherent state when the Hamiltonian is the harmonic oscillator Hamiltonian. 

For the Hamiltonian modified with deformed commutators, we see that the eigenvalues can be written as~\eqref{Eq:eigenval}
\begin{equation}
E = \hbar \omega \left( n + \nu n + \nu n^{2} + \frac{1}{2} (1 + \nu) \right) 
\end{equation}
and therefore, ignoring the constants, 
\begin{equation}
e_{n} = n \left( 1 + \nu + \nu n \right).
\end{equation}

Note that here the states $\ket{J, \gamma + \omega t}$ for different $t$ are not necessarily the same as $\ket{J, \gamma + \omega t + 2\pi}$, because $\omega$ is not related directly to the measured frequency but is merely a parameter in the Hamiltonian of Eq.~\eqref{Eq:eigenval}.

Finally, we can calculate the trajectory of the oscillator.
Using the rules of the generalised Heisenberg algebra, and the expressions for the position and momentum operators derived in the above section, we calculate the expectation value of position and momentum in this state and obtain
\begin{equation}
\begin{split}
\braket{J,\gamma | \hat{x} | J, \gamma} =&  \sqrt{\frac{2 \hbar J}{ m \omega}} \cos{\gamma}
+ \beta \sqrt{2 \hbar^{3}  m \omega }  \left\{ \frac{J^{3/2} }{4} \cos{\gamma}
\right. \\ &
\left. - \frac{J^{3/2} }{4} \cos{3 \gamma}  -  \sqrt{J} (1+J) \gamma \sin{\gamma} \right\}
\end{split}
\label{Eq:xExp}
\end{equation}
and
\begin{equation}
\begin{split}
\braket{J,\gamma | \hat{p} | J, \gamma} =&   \, - \sqrt{2 \hbar m \omega J} \sin{\gamma}
\\
& +  \beta \frac{(\hbar m \omega)^{3/2} }{2 \sqrt{2}} \left\{ J^{3/2} \sin{\gamma} +   J^{3/2} \sin{3 \gamma}
\right.
\\ &
+ 2 J^{1/2} \sin{\gamma}
- 4 \gamma \sqrt{J}(1+J)  \cos{\gamma}
\left.
\right\}.
\end{split}
\label{Eq:pExp}
\end{equation}
Since these states satisfy the relation $\e^{-i H t/\hbar} \ket{J, \gamma} = \ket{J, \gamma + \omega t}$, the time-evolved expectation values are easily obtained by replacing $\gamma$ with $\gamma + \omega t$.
In these calculations, we have assumed not only that $\beta \ll 1$ but also $\beta \gamma \ll 1$ and $\beta \omega t \ll 1$.

We choose the initial state such that the oscillator starts at rest with non-zero amplitude, i.e., $\braket{p(0)} = 0$ and $\braket{x(0)} = A$.
This condition is satisfied when $\gamma = 0$ and $J = \frac{m \omega A^{2}}{2 \hbar}$, as can be seen from Eqs.~\eqref{Eq:pExp} and~\eqref{Eq:xExp}.
Therefore, for this initial state, the expectation value of position is
\begin{equation}
\begin{split}
\braket{x(t)} =& A \cos{\omega t} + \beta \frac{m^{2} \omega^{2} A^{3}}{2} \sin{\omega t}
\\
& \times \left\{ \cos{\omega t} \sin{\omega t}  - \omega t \left(1+ \frac{2 \hbar}{m \omega A^{2}} \right)  \right\}.
\end{split}
\label{Eq:xt}
\end{equation}
The classical limit of Eq.~\eqref{Eq:xt} ($\hbar \to 0$) satisfies the low amplitude limit of the differential equation~\eqref{Eq:xt_pendulum} obtained classically, thus showing that we obtain identical results with both the methods used.

Potential future directions include connecting our formalism, which is based on the Gazeau-Klauder coherent states, with that of Ref.~\cite{Bosso2017a}, which starts with a somewhat different definition of creation and annihilation operators.
Another important open problem is to go beyond coherent states as the initial states to thermal states.
Our analysis can straightforwardly be applied to thermal states in principle, but the definition of a thermal state under the deformed commutators is unclear.

\subsection*{Data from experiment of Ref.~\cite{Smith1964a}}
\label{Sec:Data}
Table~\ref{Tab:ExpData} contains the data extracted from the experimental results obtained in Ref.~\cite{Smith1964a}.
This data was extracted by magnifying from Fig.~3 of the paper, setting the figure as a plot background and plotting markers over the figure such that the plot points coincide exactly with the points in the background.
The coordinates of the plotted points were recorded as the extracted data.

\begin{table}\centering
\renewcommand{\arraystretch}{1.2}
\begin{tabular}{c c c c c}
\toprule
$A^{2} (\mathrm{cm}^{2})$ & $T_{2 \pi} (\mathrm{s})$ &~\hspace*{0.5cm}& $A^{2} (\mathrm{cm}^{2})$ & $T_{2 \pi} (\mathrm{s})$\\
\midrule
43 & 3.47315 &~\hspace*{0.5cm}& 709 & 3.47468 \\
52 & 3.47308 &~\hspace*{0.5cm}& 837 & 3.47498 \\
132 & 3.47341 &~\hspace*{0.5cm}& 1020 & 3.47538\\
168 & 3.47342 &~\hspace*{0.5cm}& 1228 & 3.47583 \\
204 & 3.47351 &~\hspace*{0.5cm}& 1404 & 3.47633 \\
244 & 3.47363 &~\hspace*{0.5cm}& 1760 & 3.47705 \\
293 & 3.47373 &~\hspace*{0.5cm}& 1850 & 3.47736 \\
360 & 3.47396 &~\hspace*{0.5cm}& 2115 & 3.47801 \\
387 & 3.47394 &~\hspace*{0.5cm}& 2160 & 3.47798 \\
443 & 3.47409 &~\hspace*{0.5cm}& 2295 & 3.47847 \\
578 & 3.47438 &~\hspace*{0.5cm}& & \\
\bottomrule
\end{tabular}
\caption{Measured data of the time-period of a pendulum as a function of its amplitude extracted from Ref.~\cite{Smith1964a}.}
\label{Tab:ExpData}
\end{table}

\section*{Data availability statement}
All data generated or analysed during this study are extracted from Ref.~\cite{Smith1964a} and are available in this article in Table~\ref{Tab:ExpData}.

\section*{Acknowledgements}
This work was supported by the ERC Synergy grant BioQ. We gratefully acknowledge illuminating discussion 
with Gerold Brackenhofer on the measurement of pendulum oscillations and especially for bringing Ref.~\cite{Smith1964a} to our attention. We are also grateful to Shai Machnes for making available the 
QLib Mathematica package which was used in simplifying the quantum mechanical calculations. We thank Sandro Donadi, Igor Pikovski and David Vitali for helpful comments and discussions. 

\section*{Author Contributions}
M.B.P. conceived the idea and supervised the project and S.P.K performed calculations and drafted the
manuscript. Both authors discussed the results and edited the manuscript.

\section*{Competing interests}
The authors declare no competing interests.

\section*{References}

\bibliography{Notes_pendulum}

\appendix

\end{document}